\documentstyle[12pt]{article}
\newcommand{\be}{\begin{equation}}
\newcommand{\ee}{\end{equation}}
\newcommand{\bea}{\begin{eqnarray}}
\newcommand{\eea}{\end{eqnarray}}
\newcommand{\ba}{\begin{array}{l}}
\newcommand{\ea}{\end{array}}

\newcommand{\bb}{\begin{thebibliography}}
\newcommand{\eb}{\end{thebibliography}}
\newcommand{\ci}[1]{\cite{#1}}
\newcommand{\lab}[1]{\label{#1}}
\newcommand{\re}[1]{(\ref{#1})}

\newcommand{\half}{{\textstyle{\frac{1}{2}}}}
\newcommand{\cM}{{\cal M}}

\begin{document}
\begin{center} \large \bf {The axial anomaly and the conversion of gluons \\ into  photons} \footnote{Talk given at the Lake Louise Winter Institute,
18-24 February 1996}
\end{center}
\begin{center} \large
 M.M. Musakhanov$^{1,2}$,
F. C. Khanna $^{2}$
\end{center}
\begin{center}
\begin{description}
\item [\tt $1$]
Theoretical  Physics Dept, Tashkent State University, 700095,
 Uzbekistan
\item [\tt $2$]
  Department of  Physics, University of Alberta, Edmonton, Canada
 $T6G 2J1$ \\
and TRIUMF, 4004 Wesbrook Mall, Vancouver, BC,
 Canada, $V6T 2A3 $ \\
 E-mail: yousuf@phys.ualberta.ca, khanna@phys.ualberta.ca
\end{description}
\end{center}
\centerline {\bf Abstract }
The axial anomaly in the divergence of the singlet axial current in 
QCD $+$ QED leads to   low-energy  theorems for  the matrix element of this operator equation over vacuum and two--photon states and for the
matrix element  over vacuum and two--gluon states.
The solution of these theorems is related only to  the nonperturbative phenomena. 
These matrix elements are calculated in instanton vacuum generated 
N-JL type quark model  for  arbitrary $N_f .$
It is shown that this model does satisfy the   low-energy   theorems.
\\ \\
{\bf  1. Introduction.}

The axial anomaly leads to many interesting nonperturbative phenomena in physics. Among them are $B$-violation processes in electroweak $(EW)$ physics, $U_A (1)$ problem in QCD etc.  The solution of these problems  is intimately related to the  topologically nontrivial structure of the vacuum in the gauge theories. 

In this paper we apply the axial anomaly low-energy theorems  \ci{Shi} to test the chiral quark model \ci{DP86} which is based on the instanton model of QCD vacuum \ci{Shu82}.
We find that this model does satisfy these theorems.  This conclusion provides solid background to calculate the different amplitudes of nonperturbative conversion of  gluons  into hadrons and photons.       
\\ \\
{\bf  2. The low-energy theorems and axial anomaly.}

The axial anomaly in the divergence of the singlet axial current in 
QCD $+$ QED leads to a  low-energy  theorem for  the matrix elements of this operator equation over vacuum and two--photon states:
\be
 \langle 0|N_f \frac{g^2}{32\pi^2}G\tilde G | 2\gamma \rangle =
N_c \frac{e^2}{8\pi^2} \sum_{f}{Q^{2}_{f}} F^{(1)}\tilde F^{(2)}
\lab{theorem}
\ee
at $q^2 = 0$ \ci{Shi}.
Here,  $N_f$ is the number of the flavors,  $g$ is QCD coupling constant with  $2G\tilde G= \epsilon^{\mu\nu\lambda\sigma}
 G^{a}_{\mu\nu}  G^{a}_{\lambda\sigma}$ and $G^{a}_{\mu\nu}$ being the  operator of the gluon field strengths , $N_c$ is the number of the colors,
 $e$ and $Q_{f}$ are QED coupling constant and the electric charges of the quarks, respectively,
$2F^{(1)}\tilde F^{(2)} = \epsilon^{\mu\nu\lambda\sigma}
 F^{(1)}_{\mu\nu}  F^{(2)}_{\lambda\sigma},$  
$F^{(i)}_{\mu\nu} = \epsilon^{(i)}_\mu q_{i\nu} - \epsilon^{(i)}_\nu q_{i\mu},$  $\epsilon^{(1,2)}_\mu$, $q_{1,2}$ are polarizations and momenta of  photons respectively and $q=q_{1} + q_{2}$. This relation is a consequence of the absence of a massless singlet pseudoscalar boson.  Here the contribution of the quark masses is neglected. 

So, the problem is reduced to the calculation of the matrix element:

\be
 \langle 0|{g^2}G\tilde G| 2\gamma \rangle =  \epsilon^{(1)}_\mu  \epsilon^{(2)}_\nu \int \langle 0| T({g^2}G\tilde G j^{em}_\mu (x_1 )
j^{em}_\nu (x_2 ) |0 \rangle  \exp i(q_1 x_1 + q_2 x_2 ) dx_1 dx_2.
\lab{me}
\ee

As it is evident, gluons can interact with photons only through quark loops.
In perturbation theory it  leads to at least $\sim g^{4}$ result for the left hand side of  Eq. \re{theorem}
(see  e.g. recent discussion of the higher-loop contributions to the
axial anomaly \ci{FMP}).  So, the solution of this theorem is related only to  the nonperturbative phenomena connected with the structure of QCD vacuum. 

Another nontrivial low-energy theorem concerns the matrix element over
vacuum and two-gluon states:
\be
 \langle 0| g^2 G\tilde G | 2 gluons \rangle = 0
\lab{theorem1}
\ee
at the  limit $q^2 = 0.$ 

These matrix elements are calculated in the instanton vacuum generated 
N-JL type quark model \ci{DP86}, \ci{DPW95} for  arbitrary $N_f .$
\\ \\
{\bf 3. The instanton vacuum of QCD.}

The instanton is the solution of  gluodynamics in the Euclidian space\ci{BPST75}:

\be
A_{\mu}^{Ia}(x) = 2g^{-1}O_{I}^{ab}\bar\eta_{\mu\alpha}^{b} 
\frac{\rho^2 (x-z)_\nu }{\left[(x-z)^2 + \rho^2 \right] (x-z)^2 },
\lab{instantonx}
\ee
and in the momentum representation and at small $k^2$ it is:
\be
A_{\mu}^{Ia}(k) = O_{I}^{ab}\bar\eta_{\mu\alpha}^{b} 
\frac{{\rm i}4\pi^2\rho^2 k_\nu }{g k^2}(1 + O(k^2 \rho^2 )).
\lab{instantonk}
\ee
The anti-instanton solution has the same form as in Eqs. \re{instantonx} and \re{instantonk} but with $\eta_{\mu\alpha}^{b}$ instead of $\bar\eta_{\mu\alpha}^{b}$.
Here $O_{I(\bar I)} $ is the   orientation  matrix of the instanton
 $I$ (anti-instanton $\bar I$) in color space, $\bar\eta (\eta )$ -- t'Hooft factors \ci{tH76}, $\rho$ is the size and $z$ is the position of the instanton.
 For large interinstanton distances $R>>\rho$ the sum of  the instantons
and the anti-instantons is also an approximate solution. The calculation of the 
action for gluon fields leads to  sum of the actions of  free instantons and classical interinstanton potential $V(R, \rho_1 , \rho_2 , O )$, where
$O=O^{T}_{1}O_{2} $ is a matrix of relative orientation. The most important part is  the
instanton-antiinstanton potential  $V_{I\bar I}$. As it is well--known, at large distances it has a form
\be
V_{I\bar I}=
\frac{32\pi^2 }{g^2 }\frac {D \rho_{1}^{2} \rho_{2}^{2}}{  R^4 }
\lab{potential}
\ee
where orientation factor $D$ is
\be
D=\bar\eta_{\mu\alpha}^{a}O^{ab}\eta_{\mu\beta}^{b}
\left(  \frac{R_{\alpha}R_{\beta}}{R^2} - \frac{1}{4} g_{\alpha\beta}\right) .
\lab{orientation}
\ee
This  resembles the dipole-dipole interaction potential and may be attractive.

The main assumption of the model concerns small distances $R \sim \rho$.
At small distances it is assumed that there is  repulsion. This assumption  is  supported
by both   phenomenological and theoretical consideration \ci{Shu82}, \ci{DP84} (see recent discussion of this "Interacting Instanton Liquid Model" in \ci{Shu95} and references therein). This  leads to a stabilization of the size and density of instantons.  
The distribution of the number of pseudoparticles should, for large 
$\langle N \rangle $, be given by 
\be
P(N) \propto \exp\left[ -\frac{b}{4} N \left(
\log\frac{N}{\langle N\rangle} -1\right)\right] , 
\hspace{2em} N_- = N_+ = N/2
\label{DN}
\ee
and the effective size distribution of $I$'s and 
$\bar I$'s coincides, and  becomes
\be
d (\rho ) = \mbox{\rm const}\times
\rho^{b - 5} \exp\left[-\frac{b - 4}{2}\, 
\frac{\rho^2}{\overline{\rho^2}}\right] .
\label{d_rho}
\ee
When fermions are included ($N_f$ flavors), the
coefficient of the beta function $b$ is 
\be
b = \frac{11}{3} N_c - \frac{2}{3} N_f .
\label{b_fermions}
\ee

The quantities $\langle N\rangle$ and $ \overline{\rho}$ include
 effects of the instantons interactions  \ci{DPW95}.
In the following calculations, for simplicity,  
all instanton sizes are considered to be  $\bar\rho$.

Both the phenomenological estimates  and variational calculations (see recent discussion in \ci{Shu95} ) lead to a mean interinstanton distance  of
$\bar R =  \left(\frac{V}{{\langle N\rangle}}\right)^{1/4} \sim 1\, fm$ and 
a mean instanton size of $\bar\rho \sim 1/3 \, fm.$ The
small packing parameter $(\bar\rho / \bar R)^{4} = 0.012$ provides a possibility for the independent averaging over positions and orientations
of instantons. 
\\ \\
{\bf 4. The chiral quark model.}

The main assumption of this model is the interpolation formula for the 
quark propagator in the single instanton field.
 It is approximated as the sum of a free propagator
and an explicit contribution of the zero mode \cite{DP86},
\be
\left(i\hat\nabla (\xi_{I(\bar I)} ) + im \right)^{-1}_{\rm 1-inst} 
\approx
(i\hat\partial )^{-1} \;\; - \;\; \frac{\Phi_\pm (x; \xi_{I(\bar I)} )
\Phi_\pm^\dagger (y; \xi_{I(\bar I)} )}{im} .
\label{prop_zero_mode}
\ee
Here, $\hat\nabla = = \hat\partial  - i g \hat A$, $\Phi_\pm (x ; \xi_{I(\bar I)})$ is the zero mode wave function of the fermion 
 in the background of one $I (\bar I )$\ci{tH76}. It  depends on the collective instanton variables $ \xi_{I(\bar I)}$ -- the size $\rho$, the position $z$ and the orientation $O$ of the instanton.

This interpolating formula should be accurate both at small momenta
($p \ll 1/\bar\rho$), where the zero mode is dominant, and 
at large momenta ($p \gg 1/\bar\rho$), where the propagator reduces to the 
free one. In the background of an 
$N_\pm$--instanton configuration  and keeping in mind the low density of the instanton media  this formula leads to the partition function of the model:
\be
 Z_N = \int D\psi D\psi^\dagger \exp  (\int d^4 x \, \psi^\dagger i \hat\partial  \psi )  \,  W_{+}^{N_+}  \, W_{-}^{N_-}, 
\lab{Z_NW}
\ee
where
\be
W_\pm = \left( - \frac{4\pi^2\bar\rho^2}{N_c} \right)^{N_f}
\int \frac{d^4 z}{V} \det J_\pm (z), 
\ee
\be
J_\pm (z)_{fg} = \int \frac {d^4 kd^4 l}{(2\pi )^8 } \exp ( -i(k - l)z)
\, F(k) F(l) \, \psi^\dagger_f (k) \half (1 \pm \gamma_5 ) \psi_g (l) ,
\lab{J_pm}
\ee
and the contribution of the current quark masses is neglected.
The form-factor $F$ is related  to the zero--mode wave function 
 in momentum space $\Phi_\pm (k; \xi_{I(\bar I)}) $ and is equal to:
\be
F(k) = - t \frac{d}{dt} \left[ I_0 (t) K_0 (t) - I_1 (t) K_1 (t)
\right] \; \rightarrow
\left\{ \ba 1 \, \, \, \,
t \rightarrow 0 \\  \frac{3}{4} t^{-3} \, \, \, \,
 t \rightarrow \infty \ea \right. 
\\
, \nonumber
\lab{F(k)}
\ee
with  $t =\frac{1}{2} k \bar\rho.$

The   formula:
\be
 (ab)^N = \int d\lambda \exp (N ln \frac {aN} {\lambda} - N +\lambda b ).
\lab{ab^N}
\ee
 (here $N >>1)$ provides the final expression for the partition function \ci{DPW95}:

\be
 Z_N = \int D\psi D\psi^\dagger \exp (- S_{eff}),
\lab{Z_NY}
\ee
where
\be
-S_{eff} =   \int  \psi^\dagger i \hat \partial  \psi  
+ Y_{+} + Y_{-} ,
\lab{S_efY}
\ee
and
\be
Y_{\pm}=  (i )^{N_f} \lambda  
\int d^4 z \, \det J_\pm (z) =  \left(\frac{2V}{N}\right)^{N_f - 1} (i M)^{N_f} 
\int d^4 z \, \det J_\pm (z) .
\lab{Y_pm}
\ee
 The  self-consistency condition at the saddle point in Eqs. \re{Z_NY}, \re{S_efY} and \re{Y_pm} leads to
\be
4 N_c V \int \frac{d^4 k}{(2\pi )^4} \frac{M^2 F^4 (k)}{M^2 F^4 (k) + k^2}
=  N .
\lab{selfconsist}
\ee
\\ \\
{\bf 5. Calculations with the low-energy theorems.}

In the quasiclassical(saddle-point) approximation any gluon operator receives its main contribution from the instanton background.
As an example, for one  instanton (anti-instanton)$I (\bar I)$,

\be
{g^2}G^2 (x) =
\frac{192 \rho^4}{\left[ \rho^2 + (x - z)^2 \right]^4} = f(x-z),
\lab{GG}
\ee
and
\be
{g^2}G\tilde G(x) \;\; = \;\; \pm f(x-z) .
\lab{GtildeG}
\ee
In the following   the operator $g^2 G\tilde G(x)$  is considered.
Given the low density of the instanton medium it is possible to neglect  
the overlap of the fields of different instantons. In that case,  
the matrix element of the gluon operator $G\tilde G(x)$ with any other 
quark operator $Q$ is:
\be \ba
\langle {g^2}G\tilde G(x) Q \rangle_N = 
Z_{N}^{-1} \int D\psi D\psi^\dagger \exp  (\int  \psi^\dagger i \hat\partial  \psi)   
\\ \\
 \times
\left( N_{+} \left( W_{G\tilde G +} (x) Q \right) W_{+}^{N_+ - 1}   W_{-}^{N_-}  +  N_{-} \left( W_{G\tilde G -} (x) Q \right) W_{+}^{N_+ }   W_{-}^{N_- - 1} \right) ,
\lab{GtildeGQ1}
\ea \ee
where
\be
W_{G\tilde G \pm} = \pm\left( - \frac{4\pi^2\bar\rho^2}{N_c} \right)^{N_f}\int d^4 z\, f(x-z) \, \det J_\pm (z) .
\lab{W_GtildeGQ}
\ee 
The application of the formula \re{ab^N} leads to 
\be
\langle {g^2}G\tilde G(x) Q \rangle_N = 
Z_{N}^{-1} \int D\psi D\psi^\dagger \exp  (- S_{eff} )   
\left( \left( Y_{G\tilde G +} (x) + Y_{G\tilde G -} (x)\right) Q\right) ,
\lab{GtildeGQ2}
\ee
where
\be
Y_{G\tilde G\pm}(x) = \pm \left(\frac{2V}{N}\right)^{N_f - 1} (i M)^{N_f} 
\int d^4 z\, f(x-z) \, \det J_\pm (z) , 
\lab{Y_GtildeGQ}
\ee
and we take $N_{\pm} = N/2$ to preserve  $CP$ invariance.
 
In this particular case, the calculation of the matrix element, Eq. \re{me}, can be reduced to the calculation of the the partition function 
\be
\hat Z_{N}  [\kappa , a] = 
Z_{N}^{-1} \int  D\psi D\psi^\dagger \exp  (- \hat S_{eff}  ) ,
\lab{hatZ_N}
\ee 
with the 
effective action, $\hat S_{eff},$ in the presence of an external electromagnetic  field, $a_\mu ,$ and an external field, $\kappa (x),$ is given as
\be
 -\hat S_{eff}  =  \int  \psi^\dagger i \hat D  \psi  
+ Y_{+} + Y_{-} + \int dx \left(Y_{G\tilde G +} (x) + Y_{G\tilde G -} (x)\right) \kappa (x),
\lab{hatS_ef1}
\ee
where $\hat D = \hat\partial  - i e Q_{f} \hat a.$

It is clear that
\be
 \langle 0| T({g^2}G\tilde G(x) j^{em}_\mu (x_1 )
j^{em}_\nu (x_2 ) |0 \rangle  = \frac {\delta \hat Z_{N}  [\kappa , a] }
{\delta \kappa (x) \delta a_\mu (x_1) \delta a_\nu (x_2) } |_{ \kappa , a = 0}
\lab{GtildeGj^em}
\ee
Finally,  Eq. \re{hatS_ef1} can be rewritten as:
\be\ba
 -\hat S_{eff}  = 
\int \psi^\dagger i\hat D \psi +  \\
\left(\frac{2V}{N}\right)^{N_f - 1}  
\int dz \, \left( 1 + \int dx \kappa (x) f(x-z)\right) \det (i MJ_{+} (z)) + \\
\left(\frac{2V}{N}\right)^{N_f - 1}  
\int dz \, \left( 1 - \int dx \kappa (x) f(x-z)\right) \det (i MJ_{-} (z)).
\lab{hatS_ef2}
\ea\ee
\\ \\
{\bf 6. Bosonization of the partition function $\hat Z_{N}  [\kappa , a].$}

Another remarkable formula 
\be
\exp (\lambda \det [i A] ) =
\int d\cM \exp\left[ - (N_f - 1) \lambda^{-\frac{1}{N_f - 1}}
(\det\cM )^{\frac{1}{N_f - 1}} + i tr (\cM A) \right] 
\lab{expA}
\ee
is used in the following discussion.
It is possible to check this  by  the saddle point approximation  of the integral.

It is convenient to introduce:
$$ M_{\pm} (z) = 
 \left( 1 \pm \int dx \kappa (x) f(x-z)\right)^{(N_f - 1)^{-1} }M$$
and rewrite
\be \ba
 -\hat S_{eff}  = 
\int \psi^\dagger i\hat D \psi + 
\\ \\
\left(\frac{2V}{N}\right)^{N_f - 1}  
\int dz \,  \det (i  M_{+} (z) J_{+} (z)) +
\left(\frac{2V}{N}\right)^{N_f - 1}  
\int dz \, \det (i  M_{-} (z) J_{-} (z)) .
\lab{hatS_ef3}
\ea \ee
By using the formula, Eq.\re{expA}, it is easy to show that:
\be\ba
\exp \left( { \left(\frac{2V}{N}\right)^{N_f - 1} \int dz \,  
\det (i  M_{\pm}(z)J_{\pm} (z)) } \right) = \\
\int D\cM_{\pm} \exp \left[ \int dz \left( - (N_f - 1) 
\left(\frac{2V}{N}\right)^{ - 1} 
(\det\cM_{\pm} )^{\frac{1}{N_f - 1}}\right) + 
i Tr (\cM_{\pm}M_{\pm}J_{\pm}) \right] 
\lab{aplexpA}
\ea\ee
By using this relation, Eq.\re{aplexpA},   the path integral over quarks in the partition function, Eq.\re{hatZ_N},   the effective action $\hat S_{eff}$, describing mesons
in the presence of the external fields $a_\mu$ and $\kappa$     is
\be\ba
 -\hat S_{eff}[ \cM_{\pm}, a, \kappa ] = \int dz \left( - (N_f - 1) 
\left(\frac{2V}{N}\right)^{ - 1} 
(\det\cM_{\pm} )^{\frac{1}{N_f - 1}}\right) + \\ \\
Tr \ln \left( i\hat D + i \cM_{+}MF^2 \left( 1 + ( \kappa  f)\right)^{N_{f}^{-1}} \frac {1}{2}(1+\gamma_5 )  +
 i \cM_{-}MF^2
\left( 1 -  (\kappa f)\right)^{N_{f}^{-1}} \frac {1}{2}(1- \gamma_5 ) \right)
\lab{hat S_ef4}
\ea\ee
For the processes without mesons the partition function is:
\be\ba
\hat Z_{N} [\kappa , a]  = \nonumber \\ \\
\exp Tr \ln \left( i\hat D + iMF^2 \left( 1 + ( \kappa  f)\right)^{N_{f}^{-1}} \frac {1}{2}(1+\gamma_5 )  +
 iMF^2
\left( 1 -  (\kappa f)\right)^{N_{f}^{-1}} \frac {1}{2}(1- \gamma_5 ) \right)       
\nonumber \\  \\
\times \left( i \hat \partial  +  iMF^2 \right)^{-1},
\lab{hatZ_N2}
\ea\ee
where $( \kappa  f) =  \int dx \kappa (x) f(x-z) .$

{\bf  7. The low-energy theorem for the matrix element between vacuum and two-photons states. }

The matrix element,  Eq. \re{me},  is generated by  
$$\frac {\delta \hat Z_{N}  [\kappa , a] }
{\delta \kappa (x) \delta a_\mu (x_1) \delta a_\nu (x_2) } |_{ \kappa , a = 0}
 $$
 and is given by the Feynman diagram,  Fig.1:

\let\picnaturalsize=N
\def\picsize{1.5in}
\def\picfilename{conversion1.eps}
\ifx\nopictures Y\else{\ifx\epsfloaded Y\else\input epsf \fi
\let\epsfloaded=Y
\centerline{\ifx\picnaturalsize N\epsfxsize \picsize\fi \epsfbox{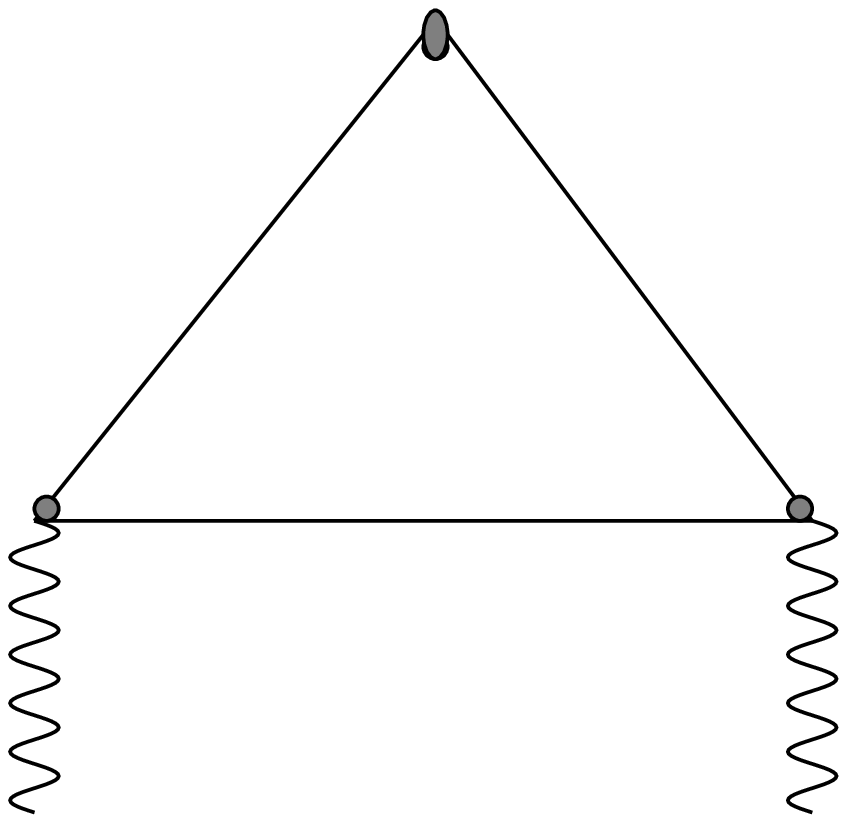}}}\fi
\centerline{Fig. 1}

As it is clear from Eq. \re{hatZ_N2} the factors in the vertices in the diagram are $eQ_{f}\gamma_{\mu}$ and 
$i M f F^{2}\gamma_{5}N_{f}^{-1}$.

We must calculate ${\Delta (q^2 )}$ (in the 
$ \lim {q^2\rightarrow 0}$), which is defined by:
\be \ba
(2\pi )^4 \delta (q - q_1 - q_2 ) \Delta (q^2 ) = 
\int dx \exp (-iqx)
 \langle 0|{g^2}G\tilde G(x)| 2\gamma \rangle = \\ \\
e^{2} \epsilon^{(1)}_\mu  \epsilon^{(2)}_\nu \int \langle 0| T({g^2}G\tilde G(x) j^{em}_\mu (x_1 )
j^{em}_\nu (x_2 ) |0 \rangle \exp i(-iqx + q_1 x_1 + q_2 x_2 ) dx dx_1 dx_2.
\lab{Delta1}
\ea \ee
It is clear  from  the previous consideration that
$$\Delta (q^2 ) = \epsilon^{(1)}_\mu  \epsilon^{(2)}_\nu   f(q^2 )
N_c  e^2 \sum_{f}{Q^{2}_{f}} $$
 $$\times Tr [ \int {{d^4 p \over \left( 2\pi  \right)^4}
\frac {iM F_{1}F_{2} \gamma_5 
( \hat p - \hat q_1 + iM F^{2}_{1} ) 
\gamma_{\mu}
(\hat p + iM F^{2}_{3} )
\gamma_{\nu}
(\hat p + \hat q_2 + iM F^{2}_{2} ) }
{ (  (p - q_1 )^2 +  M^2  F_{1}^{4}) 
(  p^2 +  M^2  F^{4}_{3} )
(  (p + q_{2} )^{2} +   M^{2}  F_{2}^{4} ) }} $$
\be 
+ ( \mu \rightleftharpoons \nu , \, q_1 \rightleftharpoons q_2 )].
\lab{Delta2}
\ee
Here $F_{1}=F((p-q_1 )$, $F_{2}=F(p+q_2 )$, $F_{3}=F(p)$ and
\be
f(q^2 )=\int dx \exp(-iqx) f(x)
\lab{fq^2}
\ee
 is the form-factor of the one-instanton
contribution to ${g^2}G\tilde G$. At $q^2 = 0$  
$$f(q^2 =0)= 32\pi^2 .$$
It is easy to show that the trace in Eq.\re{Delta2} can be reduced to
$$
8M^2 \epsilon^{\mu\nu\lambda\sigma}q_{1\lambda}q_{2\sigma}
\Gamma (q^2 ),
$$
where
\be
\Gamma (q^2 ) = 
  \int {{d^4 p \over {\left( 2\pi^4 \right)}}
{  F_{1}F_{2} F_{3}^{2}\over {\left(  (p - q_1 )^2 +  M^2  F^{4}_{1} \right) 
\left(  p^2 +  M^2  F^{4}_{3} \right)
\left(  (p + q_2 )^2 +  M^2  F^{4}_{2} \right)}}} + ...
\lab{Gamma}
\ee
It is possible to calculate  this integral if we put $F=1.$
In this approximation 
\be
\Gamma (q^2 ) = {1 \over {16\pi^2 q^2 }}
 \int_{0}^{1}{{dx \over {1-x}} ln \left( 1 + {x(1-x)q^2 \over M^2}\right)} .
\lab{Gamma1}
\ee
At  a small values of $q^{2}, $  we have
\be 
\Gamma (q^2 ) =  {1 \over 32\pi^2 M^2 }
\left( 1 - {q^2 \over 12M^2}\right)
\lab{Gamma2}
\ee
As a result,  the left side of the low-energy theorem, Eq.\re{theorem},  is
\be
(N_f \frac{g^2}{32\pi^2})(\frac {4e^2 N_c}{g^2 N_f} \sum_{f}{Q^{2}_{f}} )
F^{(1)}\tilde F^{(2)}
\lab{theorem2}
\ee
and this coincides with the right side of  Eq.\re{theorem}.
\\ \\
{\bf  8. The low-energy theorem for the matrix element between vacuum and two-gluons states.}

Here we present, without details,  calculations related to Eq. \re{theorem1}.
This matrix element can be written in the form:
\be\ba
\langle 0| g^2 G\tilde G | g(\epsilon^{(1)} , q_1 ), g(\epsilon^{(2)} , q_2 ) \rangle  = 
 \epsilon^{(1)a_1}_{\mu_1}  \epsilon^{(2)a_2}_{\mu_2}  \\ 
\lab{me2} \\
\times\int  \partial^{2}_{2} \,\partial^{2}_{1}\, \langle 0| T{g^2}G\tilde G A_{\mu_1}^{a_1}(x_1 ) A_{\mu_2}^{a_2}(x_2)
|0 \rangle  \exp i(q_1 x_1 + q_2 x_2 ) dx_1 dx_2.
\ea\ee
Here $A_{\mu}^{a}(x )$ is a total gluon field, 
$\epsilon^{(i)a_i}_{\mu_i} , q_i $ are
the polarization and the momentum of gluons respectively.

As usual, we expand the total field $A_{\mu}^{a}(x )$ around  the instanton background. The main term in Eq.\re{me2} is the contribution of the instanton background and is order $\sim O(g^{-2}).$ The next term is the
contribution of the perturbative fluctuations over instanton background and is order  $\sim O(g^{2}).$ 
It is easy to see from previous considerations that  $ O(g^{-2})$
term is given by the formula 

\be
Z_{N}^{-1} \int D\psi D\psi^\dagger \exp  (- S_{eff} )   
\left( \left( Y_{G\tilde GAA +} (x) + Y_{G\tilde GAA -} (x)\right) Q\right) ,
\lab{GtildeGAAQ2}
\ee
where
\be\ba
Y_{G\tilde GAA \pm} = \pm \left(\frac{2V}{N}\right)^{N_f - 1} (i M)^{N_f} 
\int d^4 z\, f(x-z) \\
\lab{Y_GtildeGAAQ} \\
  \times \int dO
(-\partial^{2}_{1})A_{\mu_1}^{I(\bar I)a_1}(x_1)
(-\partial^{2}_{i})A_{\mu_2}^{I(\bar I)a_2}(x_2)
 \det J_\pm (z) , 
\ea\ee
Here the instanton(anti-instanton) is located at the point $z$ with its orientation $O$. 

Repeating of the bosonization trick   leads to the result for the $ O(g^{-2})$ contribution  which is proportional to
$$ Tr[ (i\hat\partial + iMF^2)^{-1}iM F^2 \gamma_5 ].$$
It is clear that this $Tr$  and as a consequence $O(g^{-2})$ term are equal to zero.

The next   $ O(g^{2})$ term is the contribution of the diagrams, Fig.2.

\let\picnaturalsize=N
\def\picsize{1.5in}
\def\picfilename{conv-fig2.eps}
\ifx\nopictures Y\else{\ifx\epsfloaded Y\else\input epsf \fi
\let\epsfloaded=Y
\centerline{\ifx\picnaturalsize N\epsfxsize \picsize\fi \epsfbox{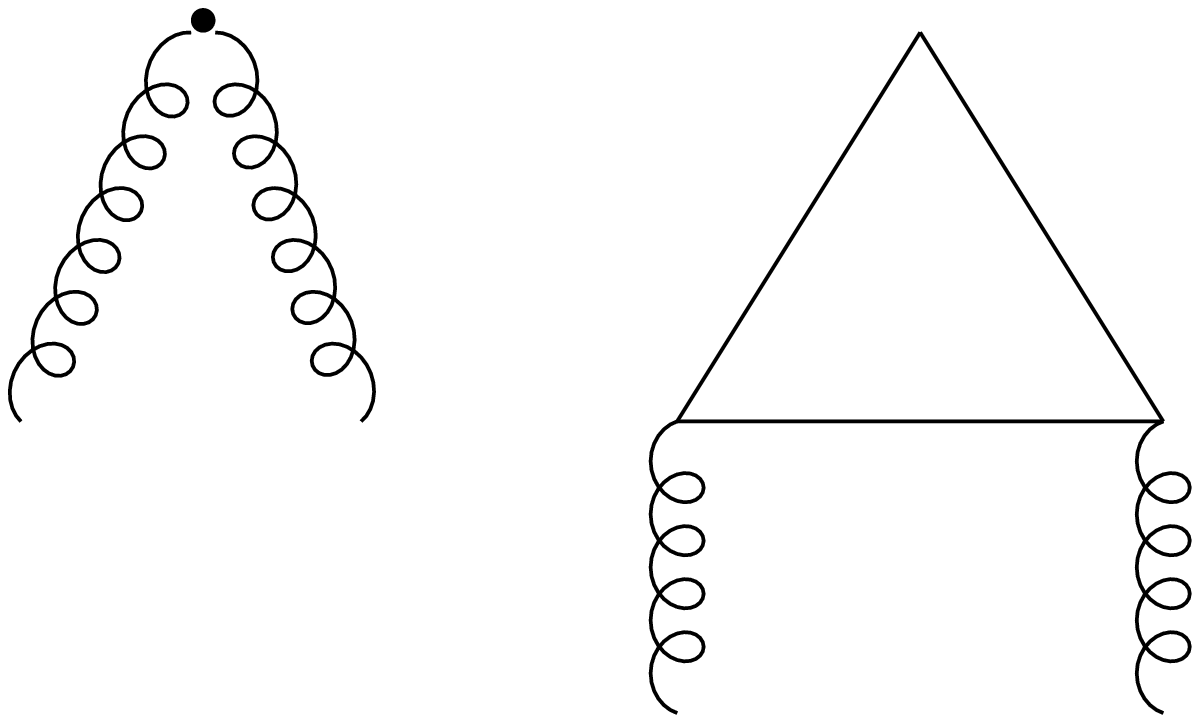}}}\fi
\centerline{Fig. 2}

It is clear that the first diagram is the direct contribution of  the operator
$g^2 G\tilde G$ which is equal to
$$- g^2 G^{(1)} \tilde G^{(2)},$$
where $2G^{(1)} \tilde G^{(2)} = \epsilon^{\mu\nu\lambda\sigma}
 G^{(1)a}_{\mu\nu} G^{(2)a}_{\lambda\sigma},$  $G^{(i)a}_{\mu\nu}=
\epsilon^{(i)}_{\mu}  q_{i\nu} - \epsilon^{(i)}_{\nu}  q_{i\mu}.$
 
The factors in the vertices of the second loop--diagram are 
$g \lambda_{a} /2 \gamma_{\mu}$ and 
$i M f F^{2}\gamma_{5}N_{f}^{-1}$. An account of the contribution from all flavors gives the coefficient $N_{f}.$ 

A comparison with the  previous calculations  (Eq.\re{Delta1},
\re{Delta2}) leads to the result that the contribution of the second loop--diagram is
equal in magnitude to the contribution of the first diagram at $q^2 = 0$ but
of opposite in sign.

Then,  terms of $ O(g^{2})$ are equal to zero in the  limit $q^2 \rightarrow 0$. 
\\ \\
{\bf  9. Conclusion.}
   
Thus we conclude that the instanton vacuum generated chiral quark model  satisfies the low-energy theorems,  Eq.\re{theorem} and \re{theorem1}.
This  provides solid background to calculate the different amplitude of nonperturbative conversion of  gluons  into hadrons and photons.  

We are planning to apply this  approach to the transitions   between the states in heavy quarks systems and to the  $\gamma  \gamma$ collision processes. 
\\ \\
{\bf Acknowledgements.}

The  work of M. Musakhanov is supported in part  by the grant INTAS-93-0239. The work of F.Khanna is
supported in part by the Natural Sciences and Engineering research Council of Canada.

\end{document}